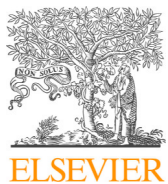
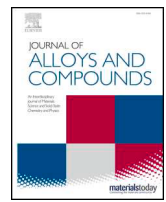

# Novel one-pot sol-gel synthesis route of Fe₃C/few-layered graphene core/shell nanoparticles embedded in a carbon matrix

Alberto Castellano-Soria [a,b,*], Jesús López-Sánchez [a,c,d], Cecilia Granados-Miralles [e], María Varela [f], Elena Navarro [a,b], César González [a,b], Pilar Marín [a,b]

[a] Instituto de Magnetismo Aplicado (IMA-UCM-ADIF), 28230 Madrid, Spain
[b] Departamento de Física de Materiales, Universidad Complutense de Madrid (UCM), 28040 Madrid, Spain
[c] SpLine, Spanish CRG BM25 Beamline, ESRF-The European Synchrotron, 38000 Grenoble, France
[d] Instituto de Ciencia de Materiales de Madrid (ICMM-CSIC), 28049 Madrid, Spain
[e] Instituto de Cerámica y Vidrio (ICV-CSIC), 28049 Madrid, Spain
[f] Instituto Pluridisciplinar & Departamento de Física de Materiales, Universidad Complutense de Madrid (UCM), 28040 Madrid, Spain

## ARTICLE INFO

## ABSTRACT

Fe₃C/few-layered graphene core/shell nanoparticles embedded in a carbon matrix are synthesized by a novel two-step surfactant sol-gel strategy, where the processes of hydrolysis, polycondensation and drying take place in a one-pot. The present approach is based on the combined action of oleic acid and oleylamine, which act sterically on the precursor micelles when a densification temperature is performed in a reducing atmosphere. The structural and magnetic evolution of the formed compounds is investigated, ranging from iron oxides such as Fe₃O₄ and FeO, to the formation of pure Fe₃C/C samples from 700 ℃ onwards. Interestingly, Fe₃C nanoparticles with a size of ~20 nm crystallize immersed in the carbon matrix and the surrounding environment forms an oriented encapsulation built by few-layered graphene. The nanostructures show a saturation magnetization of ~43 emu/g and a moderate coercivity of ~500 Oe. Thereby, an innovative chemical route to produce single phase Fe₃C nanoparticles is described, and an effective method of few-layered graphene passivation is proposed, yielding a product with a high magnetic response and high chemical stability against environmental corrosion.

© 2022 The Authors. Published by Elsevier B.V.
CC_BY_NC_ND_4.0

## 1. Introduction

The advances and developments of magnetic materials have recently focused their attention on magnetic nanocomposites based on carbides and carbon-metal alloys. Commonly such ceramic materials have exhibit noticeable hardness and high chemical resistance [1,2]. Since the discovery of such "nobility", provided by its electronic structure properties, they have been investigated for their unique catalytic [3], electrocatalytic [4,5] and remarkable good magnetic properties [6], particularly at the nanoscale [7–9]. Specifically, the nano-intermetallic iron carbide (Fe₃C) compounds are being broadly applied in energy and biomedical-related fields due to the magnetic and chemical activity of iron, and/or mechanical strength and chemical inertness of carbon [10], establishing a good protection against chemical corrosion or oxidation [11]. In terms of magnetic properties, Fe₃C is a ferromagnetic material at room temperature and displays a saturation magnetization (Mₛ) of ~140 emu/g [12]. Thus, having a Mₛ degradation value regarding the α-Fe (217 emu/g) [13] not as highlighted as those exhibited from Fe₃O₄ (92–100 emu/g) [14], α-Fe₂O₃ (0.3–1 emu/g) [14,15], γ-Fe₂O₃ (60–80 emu/g) [14], and/or ε-Fe₂O₃ (15–20 emu/g) [16,17]. Therefore, if the magnetic properties are combined with its structural and thermodynamic properties, Fe₃C is considered as a promising material for multiple technological applications requiring a high magnetic response with high chemical stability.

A high chemical stability extends the lifetime of the nanoparticles (NPs) and reduces their toxicity in physiological environments, making them suitable for application in biological media: as innovative magnetically controlled nanoplatforms for many prospective biomedical applications [18], drug and gene delivery systems, detection of diseases, hyperthermia [19,20], biochemical sensing, magnetic resonance imaging [21], even in other fields like electromagnetic microwave absorption [22,23], or spintronics [24].

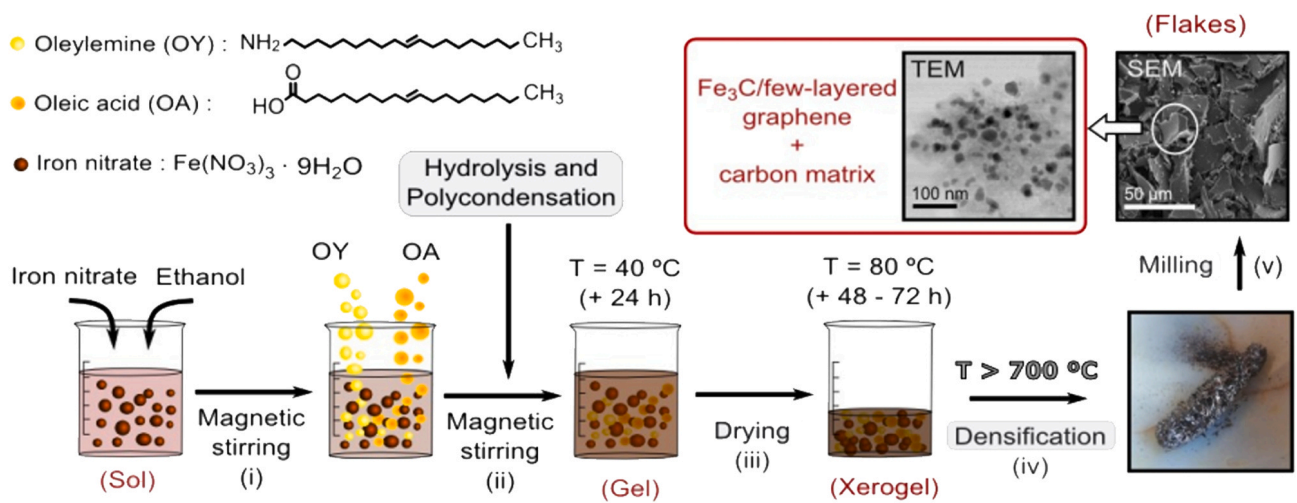

**Fig. 1.** Sol-gel synthesis scheme to obtain Fe₃C/few-layered graphene core/shell nanostructures. (i) Iron nitrate nonahydrate salt precursor is diluted in absolute ethanol and the solution is kept under magnetic stirring for 15 min (ii) OY is added dropwise into the dissolution and after 15 min, oleic acid is incorporated in the same way into the solution. The solution is kept for 24 h at 40 °C under a vigorous magnetic stirring. (iii) Drying step for 48 h at 80 °C. (iv) Densification treatment performed between 500 and 800 °C in a nitrogen atmosphere showing the lower bound (T > 700 °C) for the consolidation of Fe₃C. (v) Milling of the resulting powder on an agate mortar. The obtained flakes (scanning electron microscopy (SEM) image) composed by small NPs (transmission electron microscopy (TEM) image).

Although this material has already spawned a wide field of research involving a large and assorted applications in the steel-based industry [25], there are still no reproducible synthesis methods to obtain pure Fe₃C NPs with a control of particle size and physical properties, which might be partly due to the elevated temperature and reducing conditions required for their synthesis. In the recent literature can be found some successful Fe₃C NPs synthesis routes, for example based on nanocasting [7], metal-organic chemical vapor deposition [26], in-flight plasma treatment of gas aggregation source of NPs [27], colloidal synthesis [28], and detonation decomposition reaction [29]. Lately, the traditional sol-gel method for the synthesis of metal oxides NPs [30] has turned around in some different variants that has become one of the most promising tools for its industrial production. Sol-gel technique provides high versatility in the use of precursors, the morphology can be controlled with relatively narrow size distributions, and chemical reactions can take place at room temperature [31,32]. Advances commonly draw from a starting homogeneous gel-like-network made of organic gelators ("urea-glass-route") or using biopolymers like gelatin, followed by the formation of Fe₃C NPs by xerogel carbothermal reduction, under inert atmosphere, of prior nucleation of Fe₃O₄ into Fe₃C-carbon NPs [8,33,34]. In this framework, as a variant of "urea-glass-route", the oleic acid (OA) and oleylamine (OY) macromolecules have shown their capacity for acting as surfactants in sol-gel synthesis for nickel-carbon NPs by a calcination process at 320 °C in a nitrogen atmosphere [35]. In contrast of their well-known behavior as a surfactant stabilizer for obtaining homogenous dispersive NPs [36]. In addition, a solvothermal synthesis of capped Fe₃O₄ NPs have employed OA and OY as surfactants and in-situ photoelectron spectroscopy characterization as a function of the temperature revealed a NP surface reduction with the presence of Fe₃C at 650 °C [37].

In this study, we propose a novel non-aqueous sol-gel synthesis route with OA and OY as surfactants. The role of these precursors is to act as steric agents to control the particle size and reducing agents with the aim of promoting the carbonization of the precursor micelles instead of its oxidation, advantageously, at very much lower temperatures (~700 °C) [34] than the known use of ore-coal as a carbon source (> 1000 °C) [38]. Our non-aqueous approach uses only ethanol as a solvent for precursor iron salts, against other synthesis reported in literature [34], avoiding the conventional water addition for the use of alcohol-dissolved alkoxides as precursors. Thus, no excess of water is present, promoting a fast hydrolysis and polycondensation of the precursor micelles obtaining more homogeneous and reproducible samples. The presence of an excess of carbon provides an adequate incorporation of the aliphatic carbon from the surfactants into the prior oxide NPs promoting the formation of the Fe₃C nanostructures at densification temperatures above 650 °C. In addition, the similar linear-chain structure of the surfactants makes them cooperate effectively, constraining the size of the precursor micelles. Compositional, structural, and magnetic studies are carried out as a function of the densification temperature, establishing a thermal limit for the formation of Fe₃C single phase. Furthermore, it is revealed that the Fe₃C NPs are encased by several oriented graphene layers, presenting a core/shell structure. Therefore, the two-step surfactant method promotes the formation of graphene structures surrounding the Fe₃C NPs that can provide multifunctional properties with competitive magnetic responses at room temperature.

## 2. Experimental techniques

### 2.1. Synthesis of the Fe₃C/few-layered graphene core/shell nanostructures by sol-gel

Fe₃C/few-layered graphene core/shell nanostructures were prepared using a novel sol-gel synthesis based on a two-step surfactant strategy. As a first step, 12 mmol of Fe(NO₃)₃.9 H₂O (Sigma-Aldrich +98%) were dissolved in absolute ethanol (PanReac +99%). The solution was kept under a vigorous magnetic stirring for 15 min at 40 °C (Fig. 1i). Then, 30 mmol of OY (ACROS Organics 80–90%) was added dropwise to ensure an adequate homogenization of the micelles in the solution, avoiding the formation of undesirable organic aggregates. After 15 min, 30 mmol of OA (Sigma-Aldrich 90%) were added, maintaining a molar ratio of 1:1 respectively. Finally, 100 µL of HNO₃ were added to keep the pH among 1–2, promoting the room temperature hydrolysis and polycondensation processes at that stage.

The resulting solution is magnetically stirred for 24 h (Fig. 1ii), promoting the homogenization of the sol and the evaporation of the alcoholic products. In this first stage, hydrolysis and polycondensation happens at the same time. The OY adheres to the surface of the first ionic micelles stabilizing them and the oleic acid separates the micelles formed, increasing the steric capacity of the hydrophobic tails of oleylamine itself. The temperature was subsequently set to

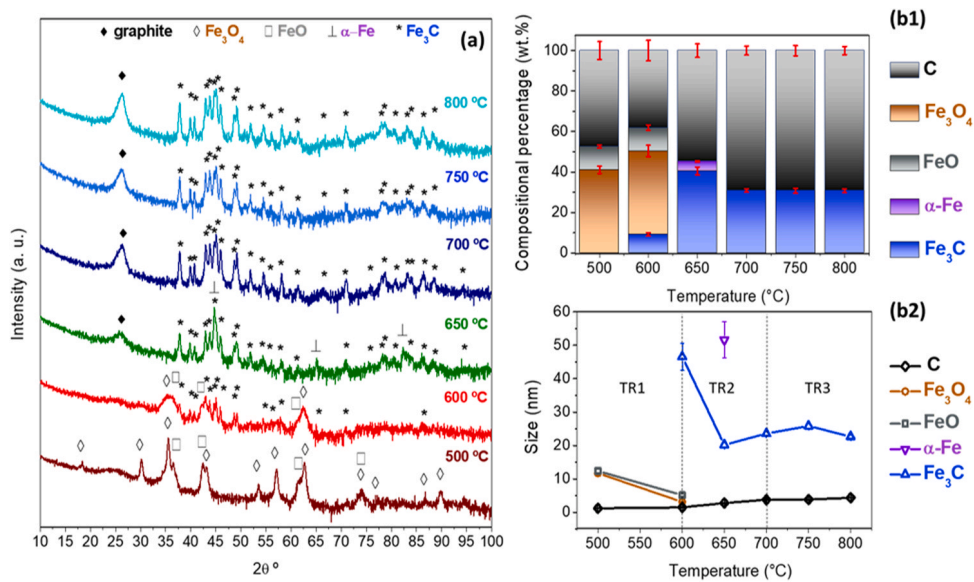

**Fig. 2.** (a) Some XRD diffractograms of samples densified between 500 and 800 °C and the contribution of each phase predicted by Rietveld refinement. (b1) Crystallite size and (b2) compositional percentage obtained from Rietveld analyses.

80 °C on a hot-plate and kept there for 48 h in an air environment, and the magnetic stirring was also maintained. From this point, the viscosity was gradually increased, and no alcoholic groups remained. The OY and OA did not evaporate at 80 °C due to their high boiling point, 364 and 360 °C respectively [39], ensuring the high carbon content required for the formation of iron carbides. In such a way, the gel aging occurred, and a dark brown xerogel was obtained (Fig. 1iii).

The densification of the xerogels was carried out in a horizontal tube furnace between 500 and 800 °C under a nitrogen flow of 20–50 cm³/min of 99.8% of purity. A heating rate of 5 °C/min was fixed and the system was kept for 1 h at the selected densification temperature (Fig. 1iv). For densification temperatures equal or greater than 650 °C, the carbothermal reduction occurs and the oxygen present is gradually removed with temperature, obtaining pure Fe₃C/few-layered graphene core/shell nanostructures. Afterwards, a cooling rate of 10 °C/min was set up to room temperature. To conclude, the resulting powder is milled manually using an agate mortar (Fig. 1v).

### 2.2. Morphological, structural, and magnetic characterization

The crystal structure, composition and volume-weighted average crystallite size of the samples were obtained from Rietveld refinements of X-ray diffraction (XRD) data measured using a PANalytical Empyrean system in a Bragg-Brentano configuration with Cu Kα radiation (λ = 1.542 Å) in the 2θ range 10–100°. The Rietveld analysis was performed using the software FullProf Suite [40]. Morphological features, crystalline structure, and particle size distribution of the nanostructures embedded in the carbon matrix were determined by scanning transmission electron microscopy (STEM) with a JEOL JEM ARM200CF with a cold field emission source, working with a CEOS aberration corrector operated at 200 kV.

The magnetic properties of the nanostructured powder were examined with a vibrating sample magnetometer (VSM) coupled to a physical property measurement system (PPMS model 6000 controller, Quantum Design). Magnetic hysteresis loops were collected at room temperature with a maximum applied magnetic field of 50 kOe.

## 3. Results and discussion

### 3.1. Structural properties and phase composition

A first aim is to investigate the evolution of the crystal structure and composition, from the first crystalline materials formed after the evaporation of the organic compounds to the stabilization of the single-phase Fe₃C/few-layered graphene-based nanostructures embedded in a carbon matrix. The existing chemical routes to produce Fe₃C use organic surfactants such as hexamethylenetetramine [41,42], melamine [43], and glucose [44]. In addition, the use of biopolymers like gelatine is an effective method to obtain pure Fe₃C samples following a sol-gel route [8]. All those chemical approaches require temperatures above 600 °C to produce the formation and subsequent evaporation of the CO₂ and the carburization of Fe is favored by carbothermal reduction [8]. Moreover, there are studies where Fe₃C is decomposed into metallic Fe and C products with increasing temperature, obtaining Fe/Fe₃C composites [34,42,43]. To unveil all those reacting mechanisms following the proposed two-step surfactant sol-gel strategy, the thermal range chosen for the densification process of the xerogels is between 500 and 800 °C.

Fig. 2a shows the evolution of the X-ray diffractograms with temperature. Rietveld refinements of the XRD data allowed extracting quantitative information of the composition and average crystalline sizes for each densification temperature (see Supporting Information). Three thermal regions (TR1, TR2 and TR3) can be distinguished attending to the resulting compositions (Fig. 2b1). The first one, TR1, with a predominance of iron oxides, encompasses from 500 to 600 °C. The diffractogram corresponding to 500 °C sample shows intense peaks associated to magnetite (Fe₃O₄, Fd3m [14,55]) at 2θ(º) = 30.1, 35.5, 43.1, 53.5, 57.0, 62.8 and wüstite (FeO, Fm3m [14,55]) at 2θ(º) = 36.6, 61.6, 74.0, with refined lattice parameters of a$_{Fe3O4}$ = 8.393(1) Å and a$_{FeO}$= 4.266(1) Å. The diagram corresponding to 600 °C sample, where a gradual carbothermal reduction of the prior Fe-oxides begins, shows a broadening of the iron oxide diffraction peaks. The rest of the phases present at this temperature, i.e., Fe₃O₄, FeO and graphite, are often found when synthesizing Fe₃C by other chemical routes [46] and they are precursors of the target material. In the present work, the formation of Fe₃C is confirmed by the detection of its most intense peaks at 2θ(º) = 37.7, 39.8, 40.7, 43.0–49.2, attributed to the orthorhombic unit cell of cementite (Fe$_3$C, Pnma [12,57]) with lattice parameters $a_{Fe3C}$ = 5.089(3) Å, $b_{Fe3C}$ = 6.744(4) Å, and $c_{Fe3C}$ = 4.526(5) Å. In turn, the crystallite sizes in the TR1 decrease with temperature from 11.7(2) nm to 3.1(1) nm and 12.4(5) nm to 5.1(4) nm for Fe$_3$O$_4$ and FeO respectively. In contrast, the Fe$_3$C phase has a greater size with a mean crystallite of 47(4) nm (Fig. 2b2). Regarding their compositional percentage (in wt%) displayed in Fig. 2b1, both iron oxides are constant in the TR1 with a 12% for FeO and a 41% for Fe$_3$O$_4$ respectively. The emergence of Fe$_3$C represents a 9% in the 600 ºC sample and it is established as the minimum temperature needed for its formation following the two-step surfactant strategy described in this work. The minor content of metastable phase FeO may be due to its lower thermal stability [48]. Rietveld analysis also establishes a graphite (2 H) phase contribution with lattice parameters $a_{2H-c}$ = 2.445(1) Å and $c_{2H-c}$ = 6.858(1) Å [50,59]. In addition, the Bragg reflections of the graphite peaks are relatively broad in the TR1, obtaining crystallite sizes of 1.2(1) and 1.5(2) nm respectively with a percentage content decreasing from 47% (500 ºC) to 38% (600 ºC). These results might also be correlated with a strong contribution of amorphous C from matrix, since there is probably a relatively high O content coming from the decomposition of the OA and the Fe$_3$C precursor micelles, before carbothermal reduction processes occur (600 ºC). Therefore, it is proposed that the Fe$_3$C formation is accompanied by the progressive decrease of Fe$_3$O$_4$ and FeO along with the decrease of the C contribution coming from the matrix.

The second thermal region, TR2, is a transition stage for the consolidation of Fe$_3$C at 650 ºC, and iron oxides are no longer detected (Fig. 2a). The 650 ºC diffractogram reflects intensity maximums belonging entirely to the Fe$_3$C peaks except several new rising peaks. The peak at 25.8º is associated with the (002) crystallographic direction of pristine graphite [50]. Additionally, a prominent peak located at 44.6º followed by 65.0º and 82.3º corresponds to α-Fe (Im3m) phase with a lattice parameter $a_{Fe}$ = 2.868(6) Å [60]. The crystallite size of the α-Fe phase is relatively large (52(5) nm, (Fig. 2b1)) with a compositional percentage of 4.7(4) % (Fig. 2b2) [43]. The occurrence of metallic Fe provides clues on the compositional evolution of the Fe$_3$C precursor micelles when is subjected to its progressive reduction and a subsequent carburization process. In other works, it is found that metallic Fe interacts with carbon nitride species to obtain Fe$_3$C and the graphitization of the C must be produced or catalyzed by the presence of Fe NPs at 650 ºC [43]. Therefore, both results are intimately correlated and lead to the stabilization of Fe$_3$C. However, interestingly, the Fe$_3$C crystalline size decreases abruptly from 47(4) nm (TR1) to 20.2(6) nm (TR2) against 1.5(2) – 2.8(2) nm for the graphite. This suggests that both, α-Fe emergence, and the increase of the graphite crystalline size might be caused by graphitization decomposition of the metastable Fe$_3$C phase at 600 ºC, temperature below its eutectic point (723 ºC) [52]. Hence, a breaking of the Fe$_3$C into smaller crystalline domain sizes at 650 ºC might be motivated by the pressure that results from the increase of the total volume sum of α-Fe and graphite phases, taking into account that some parts of surrounding Fe$_3$C environment was still not decomposed according to graphitization [12]: $Fe_3C(V_{cem}) \rightarrow 3Fe(0.91V_{cem}) + graphite(0.23V_{cem}) = 1.14V_{cem}$. Furthermore, we understand that the decomposing reduction of the iron oxide phases is activated through carbon oxidation ($2CO + O_2 = 2CO_2$), preceding the Fe$_3$C formation, in which the effective formation of graphite carries a slower kinetics (Fig. 2).

The third thermal region, TR3, goes from 700 to 800 ºC and evidences the predominance of the single-phase Fe$_3$C. It is worth mentioning that in most syntheses related to Fe$_3$C, good structural, catalytic, and magnetic properties are obtained, but there is a significant contribution from metallic Fe [26,34,42,43]. From the point of view of the purity attained, the two-step surfactant strategy method designed in this work represents a breakthrough in the synthesis of stable, single-phase Fe$_3$C where the chemical processes of hydrolysis, polycondensation, and drying take place in one pot. In addition, the stabilization of the crystalline domain size and phase content is also achieved with a size around 24(3) nm and 31(1)% respectively (Fig. 2b1–2). Furthermore, the carbon contribution represents a 69(3)% and future studies will be carried out by reducing the amount of surfactant added to reveal the minimum amount of carbon source (surfactants) necessary to form pure Fe$_3$C samples with no traces of metallic Fe. The stabilization of the crystalline domain size of the Fe$_3$C indicates that its growth could be constrained by the carbonaceous matrix crystallization acting as size stabilizer [53], obtaining a crystallite size of 23.6(5) nm at 700 ºC, 25.8(8) nm at 750 ºC, and 22.7(8) nm at 800 ºC. In addition, no α-Fe contribution predicted by Rietveld analysis remains at 700 ºC and above (Fig. 2), indicating a fully carburization into Fe$_3$C. Going above the eutectic temperature (723 ºC), no traces of decomposition of Fe$_3$C are observed by X-ray analyses and it is chemically stable up to 800 ºC. Likewise, no impurities of other crystalline carbide or nitride phases are present as usually happens in other investigations reported in literature [29,43].

TEM analyses have been carried out for the optimal densification temperature to obtain single-phase Fe$_3$C at 700 ºC. Fig. 3a1 shows a low magnification annular bright field (ABF) image where Fe$_3$C NPs are embedded in a carbon matrix. The shape of the NPs seems to be uneven, but mainly tending to be spheric-like, and its physical size ranges from 5 to 40 nm. The particle size distribution, Fig. 3a2, follows a normal fit with a mean value of 19.4(1) nm and full width high maximum (FWHM) value of 13.6(1) nm size-distribution. These values are within the same the order reported in the literature [8,26]. Interestingly, the obtained mean value is similar to the crystalline size of 23.6(5) nm calculated by Rietveld analyses (Fig. 1b1), indicating the NPs are single-crystal.

A high-resolution ABF image is presented in Fig. 3b to unveil the close environment and the structural properties of the NPs. A core-shell morphology is detected, where an effective passivation with few graphene layers surrounding the Fe$_3$C surface is achieved. These graphitic layers are formed when the amorphous carbon is dissolved in the Fe place by carbon-mediated metal diffusion [8]. In addition, in the XRD patterns, can be noticed that the graphite (002) peak shifts toward higher angles ~ 26º (see Fig. 2a and Supporting Information). The content of crystalline carbon in the carbon matrix increases with temperature as the growing trend of the 2θ = 25.8º intensity shows. This suggests that graphite crystallization could be leaded by the curvature of the graphitic shell of the NPs, i. e., a shifting (002) peak and increasing the shell thickness likewise. This trend is in a good agreement with observations in multiwalled carbon nanotubes with diameter increasing [54]. Here we consider that the close environment of the Fe$_3$C NPs might follow a similar behavior. Consequently, the shell is greatly settled at temperatures above 700 ºC, stabilizing the Fe$_3$C core and avoiding its decomposition and growth. Therefore, the NPs are chemically stable, and no undesirable crystalline phases are obtained.

The fast Fourier transformations (FFT) calculated from different areas of the core-shell nanostructure are shown in the insets 1–2 of the Fig. 3b respectively. The crystalline core shows interplanar distances of ~0.208 nm and ~0.226 nm corresponding to the crystallographic directions (102) and (002) of the orthorhombic Fe$_3$C Pnma 62. Hence, the zone axis subtracted corresponds with the [010] direction. Similarly, the interplanar distance of the shell is ~0.320 nm, matching with the (002) direction of graphite. The interface is abrupt and clear, free of secondary phases, as is showed by the ABF and high angle annular dark field HAADF images, Fig. 3c1-c2. The FFT inset of the ABF image unveils shared binding planes for a bluntly interface between the (002) graphite and (112) Fe$_3$C, with their corresponding interplanar distances of ~0.365 nm and ~0.196 nm respectively.

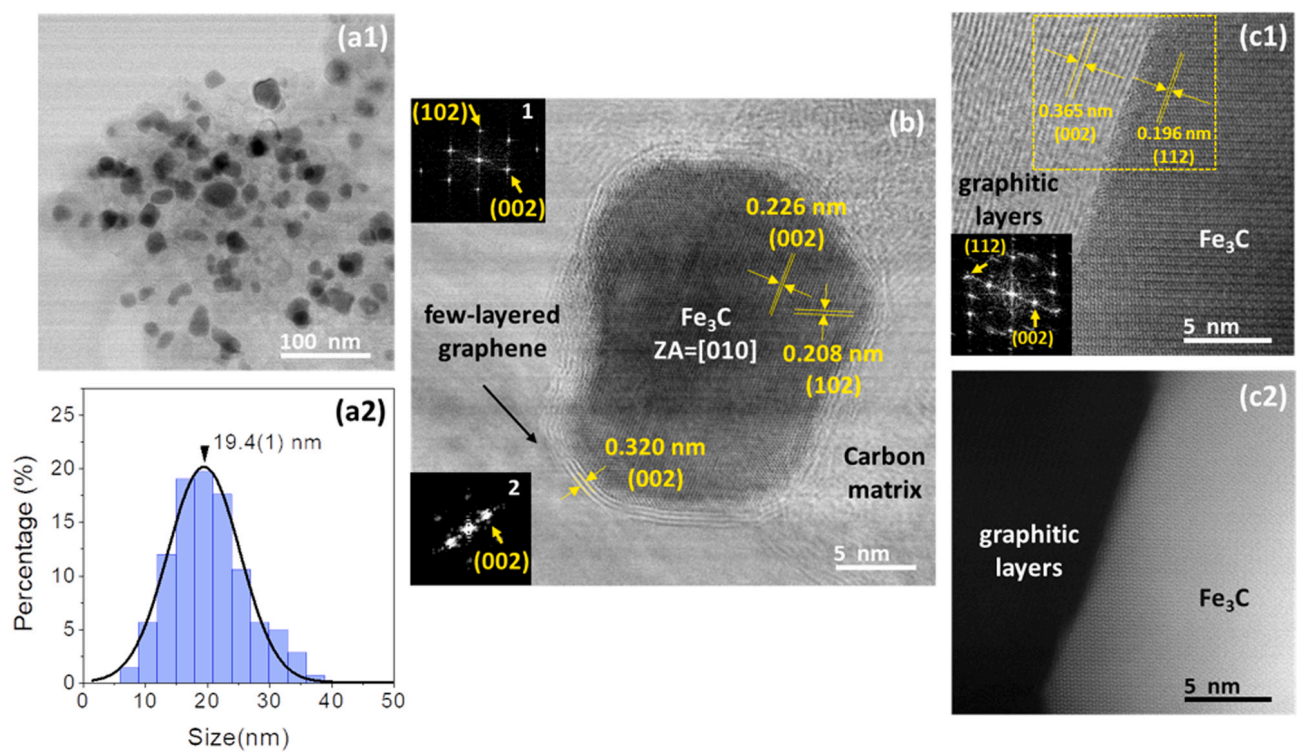

**Fig. 3.** (a1) Low magnification ABF image for Fe₃C NPs embedded in a carbon matrix and (a2) its particle size distribution. (b) ABF image of a core-shell nanostructure and insets 1/2 corresponds to the FFT of Fe₃C/few-layered graphene respectively. (c1) ABF and (c2) HAADF images for a representative interface of the core-shell nanostructure. The inset shows the FFT acquired from the yellow square region highlighted with a yellow square in Figure (c1).

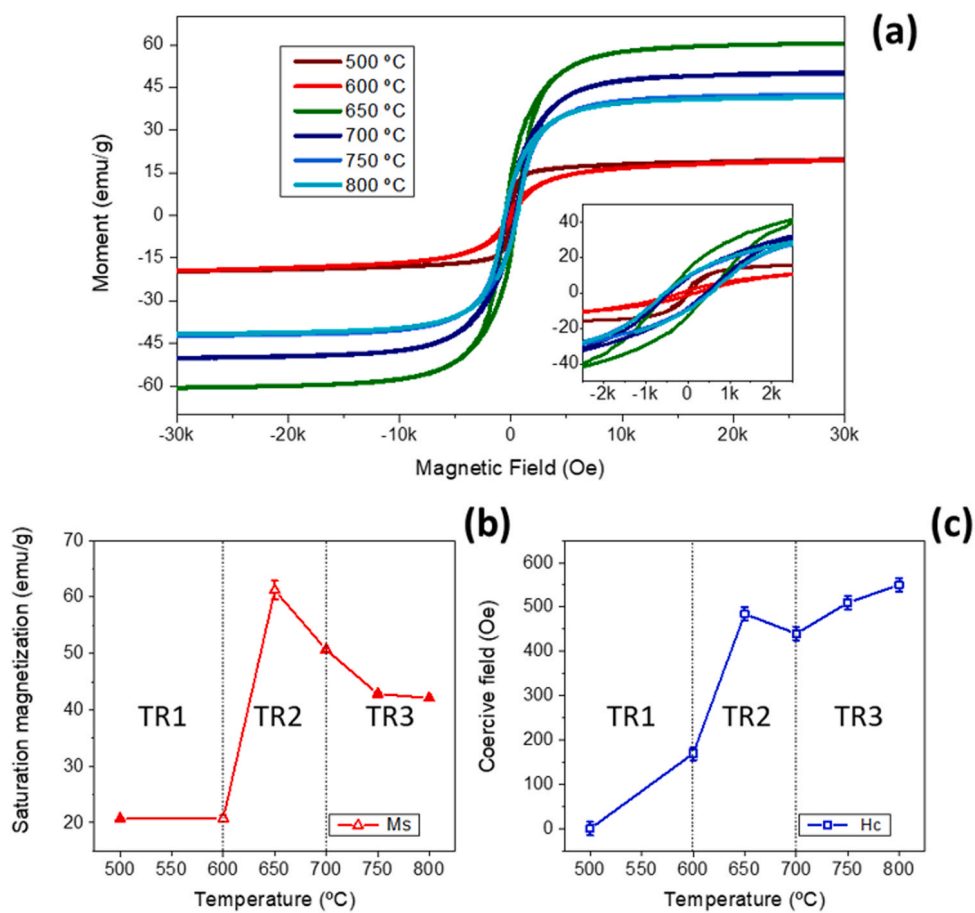

**Fig. 4.** (a) Magnetic hysteresis loops, (b) saturation magnetization, and (c) coercivity for densified samples between 500 and 800 °C.

**Table 1**

Magnetic parameters from the magnetic hysteresis loops for each densification temperature.

| Temperature (°C) | $M_s$ (emu/g) | $M_s^R$ (emu/g) | $M_r$ (emu/g) | $H_c$ (Oe) |
|---|---|---|---|---|
| 500 | 20.7 | – | < 1 | 0.5 |
| 600 | 20.8 | – | 1.3 | 170 |
| 650 | 61 | 151(8) | 12.8 | 485 |
| 700 | 50.7 | 164(5) | 8.2 | 440 |
| 750 | 42.9 | 139(5) | 8.7 | 510 |
| 800 | 42.1 | 137(4) | 9.4 | 550 |

### 3.2. Magnetic properties

Fig. 4a shows the hysteresis loops for samples densified at different temperatures. A magnetic field of 30 kOe results enough for reaching the saturation magnetization. The values of saturation magnetization ($M_s$), coercive field ($H_c$), and remanent magnetization ($M_r$) are displayed in Table 1. The three regions discussed on XRD section analysis can be differentiated in Fig. 4b for $M_s$ (T), and 4c for $H_c$ (T), displaying a good correlation. The hysteresis loops related to the TR1 present a similar $M_s$ ~21 emu/g, but the $H_c$ goes from ~0.5 to ~170 Oe with temperature. This magnetic hardening is in a good agreement with the presence of iron oxides at 500 °C and Fe₃C emergence at 600 °C. In the TR2, it is obtained the highest $M_s$ of 61(2) emu g⁻¹ and $M_r$ of 12.8(7) emu/g⁻¹, in agreement with the contribution of the α-Fe phase detected by Rietveld analysis. In terms of $H_c$, a consolidation of Fe₃C is achieved with an abrupt increase up to ~485 Oe. The values of $M_s$ obtained for the samples densified in the TR3 are 50.7(3), 42.9(3) and 42.1(2) emu g⁻¹ respectively. This trend shows a stabilization of the $M_s$ and an effective transformation from α-Fe into Fe₃C is depicted, in agreement with Rietveld analyses predictions. Those values become completely stable at 750–800 °C.

The devaluated $M_s$ values for samples respect to the Fe₃C are mainly due to the content of excess carbon from shells and matrix, which not being magnetic, both contribute to a decreasing of the $M_s$ (emu g⁻¹) by the sample mass normalization. Such $M_s$ values can be renormalized ($M_s^R$), considering: (i) the compositional percentage from Rietveld analysis in Fig. 2(b1) and (ii) an approximated $M_s$ bulk magnetic behavior for the Fe₃C NPs synthesized in this work (~20 nm), neglecting surface affects, i. e., considering Fe₃C cores hypothetically close to a $M_s$ ~140 emu/g. The $M_s^R$ for the samples belonging to TR2 and TR3, are shown in Table 1. The $M_s$ values calculated for 650 and 700 °C may would be overestimated by the α-Fe contribution not detected under the resolution of the technique or the bulk values considered to perform the estimations. Nevertheless 750 and 800 °C throws 139(5) and 137(4) emu g-1, close to bulk Fe3C value [12]. Despite being probably alike overestimated, since bulk approximation for cores neglect surface magnetic effects, this fact manifests consistent values for compositional percentages for Rietveld analysis. In such a way, the Fe₃C NPs has higher $M_s$ than predicted by taking into account the carbon matrix contribution, and they are perfectly competitive with the higher values that other authors have reported in literature [8].

Regarding the coercivity, it is constant with temperature from 650 to 800 °C with a range value of 440–550 Oe, Table 1, according with the Fe₃C nanosized value reported in literature [12].

We understand that this work constitutes the first step of this novel one-spot synthesis on the obtention of high purity Fe₃C/few-layered graphene core-shell NPs. The products obtained highlight by its purity, controlled growth through carbon matrix, its surface passivated by the few-layered shell, and a good magnetic property making them perfectly competitive and applicable for the whole vanguard proposed applications existent for this nano sized Fe₃C. In the future we hope to explore this sol-gel approach towards both the higher $M_s$ of products reducing the quantity of surfactants introduced till no contaminant phases presence like α-Fe, and explore how the ratio between OA and OY could modulate the NPs morphology.

## 4. Conclusions

In this report we have successfully stabilized a well-defined Fe₃C/few-layered graphene core/shell NPs as a single phase embedded in a carbon matrix by a novel sol-gel synthesis based on a two-step surfactant strategy, gelation, and densification, using OY and OA as surfactants. XRD data reveal the presence of FeO and Fe₃O₄ between 500 and 600 °C and a small contribution of α-Fe at 650 °C. In turn, the progressive formation of Fe₃C NPs occurs by carbothermal processes from 600 °C. The optimal densification temperature to obtain the consolidation of the single phase Fe₃C (cementite) is 700 °C and a gradual crystallization of a tridimensional carbon matrix is identified. The crystallization process of Fe₃C is due to the effective aliphatic C bonding from the surfactants with the increase of the densification temperature. The STEM analyses show that the particle size distribution is relatively homogeneous with an average particle size of 19.4(1) nm and with nearly spherical shapes. This finding is consistent with the crystallite size calculated by Rietveld analyses is 23.6(5) nm and therefore, the NPs are single crystal. Interestingly, the high-resolution STEM images unveil that the NPs are coated by oriented few-layered graphene, passivating the Fe₃C core with abrupt interfaces. The magnetic properties, $M_s$ and $H_c$, displayed are in a good agreement with their values for the crystalline phases obtained at each temperature. The $M_s$ and $H_c$ values achieved in the range 750–800 °C are ~ 43 emu/g and ~500 Oe respectively. Considering the large weight contribution of the carbon matrix of ~70%, the $M_s$ value of the Fe₃C NPs is ~138 emu/g. This value is high and close to its theoretical Fe₃C bulk value of 140 emu/g. Therefore, the two-step surfactant strategy is proposed to be an effective synthesis procedure to obtain competitive and potential magnetic properties displayed by the core/shell nanostructures embedded in a carbon matrix.

## CRediT authorship contribution statement

**Jesús López-Sánchez, Elena Navarro, Pilar Marín, César González:** Conceived the work and coordinate the research. **Alberto Castellano-Soria, Jesús López-Sánchez:** Prepared the samples. **Cecilia Granados-Miralles, María Varela, Alberto Castellano-Soria, Jesús López-Sánchez:** Carried out the structural characterization. **Alberto Castellano-Soria, Elena Navarro:** Performed the magnetic characterization. All authors wrote and revised the manuscript discussing the results and their presentation.

## Declaration of Competing Interest

The authors declare that they have no known competing financial interests or personal relationships that could have appeared to influence the work reported in this paper.

## Acknowledgments

This work has been supported by the Ministerio de Ciencia e Innovación (MCINN), Spain, through the projects: MAT2015–65445-C2–1-R, MAT2017–86450-C4–1-R, MAT2015–67557-C2–1-P, RTI2018–095856-B-C21, RTI2018–095303-A-C52, PIE: 2021–60-E-030, PIE: 2010–6-OE-013; and Comunidad de Madrid, Spain, by S2013/MIT-2850 NANOFRONTMAG and S2018/NMT-4321 NANOMAGCOST. The authors are also grateful for the electron microscopy characterization performed in the Centro Nacional de Microscopía Electrónica at the Universidad Complutense de Madrid (ICTS ELECMI, UCM), support from MCINN grant # RTI2018–097895-B-43. C.G.-M. acknowledges the financial support from MICINN through the "Juan de la Cierva" Program (FJC2018–035532-I).

## Appendix A. Supporting information

Supplementary data associated with this article can be found in the online version at doi:10.1016/j.jallcom.2022.163662.

## References

[1] C. Kral, W. Lengauer, D. Rafaja, P. Ettmayer, Critical review on the elastic properties of transition metal carbides, nitrides and carbonitrides, J. Alloy. Compd. 265 (1998) 215–233, https://doi.org/10.1016/S0925-8388(97)00297-1

[2] W.S. Williams, Physics of transition metal carbides, Mater. Sci. Eng.: A. 105–106 (1988) 1–10, https://doi.org/10.1016/0025-5416(88)90474-0

[3] J.S.J. Hargreaves, A.R. McFarlane, S. Laassiri, eds., Alternative Catalytic Materials, 2018. https://doi.org/10.1039/9781788013222.

[4] D. Li, J. Shi, C. Li, Transition-metal-based electrocatalysts as cocatalysts for photoelectrochemical water splitting: a mini review, Small 14 (2018) 1704179, https://doi.org/10.1002/SMLL.201704179

[5] Y. Liu, T.G. Kelly, J.G. Chen, W.E. Mustain, Metal carbides as alternative electrocatalyst supports, ACS Catal. 3 (2013) 1184–1194, https://doi.org/10.1021/CS4001249

[6] L.J.E. Hofer, E.M. Cohn, Saturation magnetizations of iron carbides1, J. Am. Chem. Soc. 81 (2002) 1576–1582, https://doi.org/10.1021/JA01516A016

[7] D.C. Fletcher, R. Hunter, W. Xia, G.J. Smales, B.R. Pauw, E. Blackburn, A. Kulak, H. Xin, Z. Schnepp, Scalable synthesis of dispersible iron carbide (Fe3C) nanoparticles by 'nanocasting, J. Mater. Chem. A 7 (2019) 19506–19512, https://doi.org/10.1039/C9TA06876G

[8] Z. Schnepp, S.C. Wimbush, M. Antonietti, C. Giordano, Synthesis of highly magnetic iron carbide nanoparticles via a biopolymer route, Chem. Mater. 22 (2010) 5340–5344, https://doi.org/10.1021/CM101746Z

[9] A. Meffre, B. Mehdaoui, V. Kelsen, P.F. Fazzini, J. Carrey, S. Lachaize, M. Respaud, B. Chaudret, A simple chemical route toward monodisperse iron carbide nanoparticles displaying tunable magnetic and unprecedented hyperthermia properties, Nano Lett. 12 (2012) 4722–4728, https://doi.org/10.1021/NL302160D

[10] R. Madannejad, N. Shoaie, F. Jahanpeyma, M.H. Darvishi, M. Azimzadeh, H. Javadi, Toxicity of carbon-based nanomaterials: Reviewing recent reports in medical and biological systems, Chem. -Biol. Interact. 307 (2019) 206–222, https://doi.org/10.1016/J.CBI.2019.04.036

[11] M. Messner, D.J. Walczyk, BenjaminG. Palazzo, Z.A. Norris, G. Taylor, J. Carroll, T.X. Pham, J.D. Hettinger, L. Yu, Electrochemical Oxidation of Metal Carbides in Aqueous Solutions, J. Electrochem. Soc. 165 (2018) H3107–H3114, https://doi.org/10.1149/2.0171804JES

[12] H.K.D.H. Bhadeshia, Cementite, Https://Doi.Org/10.1080/09506608.2018. 1560984. 65, 2019: 1–27. (https://doi.org/10.1080/09506608.2018.1560984).

[13] J. Crangle, G.M. Goodman, The Magnetization of Pure Iron and Nickel, 1971. (https://doi.org/10.1098/rspa.1971.0044).

[14] R.M. Cornell, U. Schwertmann, The Iron Oxides, Wiley., Weinheim; Germany, 2003. https://doi.org/10.1002/3527602097

[15] F. Bødker, M. Hansen, C. Koch, K. Lefmann, S. Mørup, Magnetic properties of hematite nanoparticles, Phys. Rev. B 61 (2000) 6826–6838, https://doi.org/10.1103/PhysRevB.61.6826

[16] J. López-Sánchez, A. Muñoz-Noval, C. Castellano, A. Serrano, A. del Campo, M. Cabero, M. Varela, M. Abuín, J. de la Figuera, J.F. Marco, G.R. Castro, O. Rodríguez de la Fuente, N. Carmona, Origin of the magnetic transition at 100 K in ε-Fe2O3 nanoparticles studied by x-ray absorption fine structure spectroscopy, J. Phys.: Condens. Matter 29 (2017) 485701, https://doi.org/10.1088/1361-648X/aa904b

[17] M. Gich, A. Roig, C. Frontera, E. Molins, J. Sort, M. Popovici, G. Chouteau, D. Martín y Marero, J. Nogús, Large coercivity and low-temperature magnetic reorientation in ε-Fe2O3 nanoparticles, J. Appl. Phys. 98 (2005) 1–5, https://doi.org/10.1063/1.1997297

[18] J. Yu, F. Chen, W. Gao, Y. Ju, X. Chu, S. Che, F. Sheng, Y. Hou, Iron carbide nanoparticles: an innovative nanoplatform for biomedical applications, Nanoscale Horiz. 2 (2017) 81–88, https://doi.org/10.1039/C6NH00173D

[19] A. Bordet, R.F. Landis, Y. Lee, G.Y. Tonga, K. Soulantica, V.M. Rotello, B. Chaudret, Water-dispersible and biocompatible iron carbide nanoparticles with high specific absorption rate, ACS Nano 13 (2019) 2870–2878, https://doi.org/10.1021/ACSNANO.8B05671

[20] M. Xing, J. Mohapatra, J. Beatty, J. Elkins, N.K. Pandey, A. Chalise, W. Chen, M. Jin, J.P. Liu, Iron-based magnetic nanoparticles for multimodal hyperthermia heating, J. Alloy. Compd. 871 (2021) 159475, https://doi.org/10.1016/J.JALLCOM.2021.159475

[21] W. Tang, Z. Zhen, C. Yang, L. Wang, T. Cowger, H. Chen, T. Todd, K. Hekmatyar, Q. Zhao, Y. Hou, J. Xie, Fe5C2 nanoparticles with high mri contrast enhancement for tumor imaging, Small 10 (2014) 1245–1249, https://doi.org/10.1002/SMLL.201303263

[22] M. Green, X. Chen, Recent progress of nanomaterials for microwave absorption, J. Mater. 5 (2019) 503–541, https://doi.org/10.1016/J.JMAT.2019.07.003

[23] R. Kumar, H.K. Choudhary, S.P. Pawar, S. Bose, B. Sahoo, Carbon encapsulated nanoscale iron/iron-carbide/graphite particles for EMI shielding and microwave absorption, Phys. Chem. Chem. Phys. 19 (2017) 23268–23279, https://doi.org/10.1039/C7CP03175K

[24] J. Szczytko, P. Osewski, M. Bystrzejewski, J. Borysiuk, A. Grabias, A. Huczko, H. Lange, A. Majhofer, A. Twardowski, J. Szczytko, P. Osewski, M. Bystrzejewski, J. Borysiuk, A. Grabias, A. Huczko, H. Lange, A. Majhofer, A. Twardowski, Carbon-Encapsulated Magnetic Nanoparticles Based on Fe, Mn, and Cr for Spintronics Applications, AcPPA 112 (2007) 305, https://doi.org/10.12693/APHYSPOLA.112.305

[25] L. Huang, R. Zhang, X. Zhou, Y. Tu, J. Jiang, Atomic interactions between Si and Mn during eutectoid transformation in high-carbon pearlitic steel, J. Appl. Phys. 126 (2019) 245102, https://doi.org/10.1063/1.5119185

[26] J. Liu, B. Yu, Q. Zhang, L. Hou, Q. Huang, C. Song, S. Wang, Y. Wu, Y. He, J. Zou, H. Huang, Synthesis and magnetic properties of Fe3C–C core–shell nanoparticles, Nanotechnology 26 (2015) 085601, https://doi.org/10.1088/0957-4484/26/8/085601

[27] H. Libenská, J. Hanuš, T. Košutová, M. Dopita, O. Kylián, M. Cieslar, A. Choukourov, H. Biederman, Plasma-based synthesis of iron carbide nanoparticles, Plasma Process. Polym. 17 (2020) 2000105, https://doi.org/10.1002/PPAP.202000105

[28] F.M. Abel, S. Pourmiri, G. Basina, V. Tzitzios, E. Devlin, G.C. Hadjipanayis, Iron carbide nanoplatelets: colloidal synthesis and characterization, Nanoscale Adv. 1 (2019) 4476–4480, https://doi.org/10.1039/C9NA00526A

[29] N. Luo, X. Li, X. Wang, H. Yan, C. Zhang, H. Wang, Synthesis and characterization of carbon-encapsulated iron/iron carbide nanoparticles by a detonation method, Carbon 48 (2010) 3858–3863, https://doi.org/10.1016/J.CARBON.2010.06.051

[30] E. S, "Traditional" sol–gel chemistry as a powerful tool for the preparation of supported metal and metal oxide catalysts, Mater. (Basel, Switz. ) 12 (2019), https://doi.org/10.3390/MA12040668

[31] J. López-Sánchez, A. Serrano, A. del Campo, M. Abuín, E. Salas-Colera, A. Muñoz-Noval, G.R. Castro, J. de la Figuera, J.F. Marco, P. Marín, N. Carmona, O. Rodríguez de la Fuente, Self-assembly of iron oxide precursor micelles driven by magnetic stirring time in sol–gel coatings, RSC Adv. 9 (2019) 17571–17580, https://doi.org/10.1039/C9RA03283E

[32] C. Giordano, M. Antonietti, Synthesis of crystalline metal nitride and metal carbide nanostructures by sol–gel chemistry, Nano Today 6 (2011) 366–380, https://doi.org/10.1016/J.NANTOD.2011.06.002

[33] C. Giordano, A. Kraupner, S.C. Wimbush, M. Antonietti, Iron carbide: an ancient advanced material, Small 6 (2010) 1859–1862, https://doi.org/10.1002/SMLL.201000437

[34] X. Wang, P. Zhang, J. Gao, X. Chen, H. Yang, Facile synthesis and magnetic properties of Fe3C/C nanoparticles via a sol–gel process, Dyes Pigments 112 (2015) 305–310, https://doi.org/10.1016/J.DYEPIG.2014.07.021

[35] P. Li, W. Jiang, F. Li, Non-aqueous sol-gel preparation of carbon-supported nickel nanoparticles, J. Sol. -Gel Sci. Technol. 65 (2013) 359–366, https://doi.org/10.1007/S10971-012-2944-Y

[36] S. Mourdikoudis, L.M. Liz-Marzán, Oleylamine in nanoparticle synthesis, Chem. Mater. 25 (2013) 1465–1476, https://doi.org/10.1021/CM4000476

[37] D. Wilson, M.A. Langell, XPS analysis of oleylamine/oleic acid capped Fe 3 O 4 nanoparticles as a function of temperature, Appl. Surf. Sci. 303 (2014) 6–13, https://doi.org/10.1016/J.APSUSC.2014.02.006

[38] K. Sun, W.-K. Lu, Mathematical Modeling of the Kinetics of Carbothermic Reduction of Iron Oxides in Ore-Coal Composite Pellets, 2008 40:1, Metall. Mater. Trans. B 40 (2009) 91–103, https://doi.org/10.1007/S11663-008-9199-6

[39] oleylamine | CAS#:112–90-3 | Chemsrc, oleic acid | CAS#:112–80-1 | Chemsrc, n. d. https://www.chemsrc.com/en/cas/112–90-3_453207.html, (https://www.chemsrc.com/en/cas/112–80-1_895631.html) (accessed 26 July 2021).

[40] J. Rodríguez-Carvajal, Recent advances in magnetic structure determination by neutron powder diffraction, Phys. B: Condens. Matter 192 (1993) 55–69, https://doi.org/10.1016/0921-4526(93)90108-I

[41] P. Zhang, X. Wang, W. Wang, X. Lei, H. Yang, Iron carbide and nitride via a flexible route: synthesis, structure and magnetic properties, RSC Adv. 5 (2015) 21670–21676, https://doi.org/10.1039/C5RA00336A

[42] P. Zhang, L. Bi, D. Zhang, X. Wang, W. Wang, X. Lei, H. Yang, Synthesis of Fe3C branches via a hexamethylenetetramine route, Mater. Res. Bull. 76 (2016) 327–331, https://doi.org/10.1016/J.MATERRESBULL.2015.12.038

[43] A. Wu, D. Liu, L. Tong, L. Yu, H. Yang, Magnetic properties of nanocrystalline Fe/Fe3C composites, CrystEngComm 13 (2011) 876–882, https://doi.org/10.1039/C0CE00328J

[44] H. Huang, M. Qin, D. Zhang, Y. Wang, Q. Wan, Q. He, B. Jia, X. Qu, Facile synthesis of sheet-like Fe/C nanocomposites by a combustion-based method, J. Alloy. Compd. C. 695 (2017) 1870–1877, https://doi.org/10.1016/J.JALLCOM.2016.11.021

[45] E. Park, O. Ostrovski, J. Zhang, S. Thomson, R. Howe, Characterization of phases formed in the iron carbide process by X-ray diffraction, mossbauer, X-ray photoelectron spectroscopy, and raman spectroscopy analyses, 2001 32:5, Metall. Mater. Trans. B 32 (2001) 839–845, https://doi.org/10.1007/S11663-001-0071-1

[46] F. Schrettle, Ch Kant, P. Lunkenheimer, F. Mayr, J. Deisenhofer, A. Loidl, W"ustite: Electric, thermodynamic and optical properties of FeO, Eur. Phys. J. B 85 (2012), https://doi.org/10.1140/epjb/e2012-30201-5

[47] D. Matatagui, J. López-Sánchez, A. Peña, A. Serrano, A. del Campo, O.R. de la Fuente, N. Carmona, E. Navarro, P. Marín, M. del Carmen Horrillo, Ultrasensitive NO2 gas sensor with insignificant NH3-interference based on a few-layered mesoporous graphene, Sens. Actuators, B: Chem. 335 (2021) 129657, https://doi.org/10.1016/J.SNB.2021.129657

[52] J. Zhang, A. Schneider, G. Inden, Cementite decomposition and coke gasification in He and H2–He gas mixtures, Corros. Sci. 46 (2004) 667–679, https://doi.org/10.1016/S0010-938X(03)00177-X

[53] C. Z, Q. M, Z. C, G. Y, J. B, Facile route for synthesis of mesoporous graphite en-capsulated iron carbide/iron nanosheet composites and their electrocatalytic activity, J. Colloid Interface Sci. 491 (2017) 55–63, https://doi.org/10.1016/J.JCIS.2016.11.086

[54] D.K. Singh, P.K. Iyer, P.K. Giri, Diameter dependence of interwall separation and strain in multiwalled carbon nanotubes probed by X-ray diffraction and Raman scattering studies, Diam. Relat. Mater. 19 (2010) 1281–1288, https://doi.org/10.1016/J.DIAMOND.2010.06.003

[55] Crystallography Open Database: Information card for entry 9008636.

[56] Crystallography Open Database: Information card for entry 1010369.

[57] Crystallography Open Database: Information card for entry 2300066.

[59] Crystallography Open Database: Information card for entry 1200017.

[60] Crystallography Open Database: Information card for entry 4113928.